\documentstyle[aps,preprint,eqsecnum]{revtex}
\begin{document}
\title{Modular Invariance, Self-Duality and
The Phase Transition Between Quantum Hall Plateaus}
\author{Eduardo Fradkin$^{*}$ and Steven Kivelson$^{\dagger}$}
\address{
Department of Physics , University of Illinois at Urbana-Champaign$^{*}$
, 1110 W.Green St, Urbana, IL 61801-3080 and 
Department of Physics, University of California Los Angeles$^{\dagger}$, 
CA 90024 
}
\maketitle
\begin{abstract}
 We investigate the problem of the superuniversality of the phase transition 
between different quantum Hall plateaus. We construct a set of models which 
give a qualitative description of this transition in a pure system of 
interacting charged particles. One of the models is manifestly invariant under 
both Duality and Periodic shifts of the statistical angle and, hence, it has a 
full Modular Invariance. We derive the transformation laws for the correlation 
functions under the modular group and use them to derive symmetry constraints 
for the conductances. These allow us to calculate exactly the conductivities at 
the modular fixed points. We show that, at least at the modular fixed points, 
the system is critical. Away from the fixed points, the behavior of the model 
is determined by extra symmetries  such as Time Reversal. We speculate that if 
the natural connection between spin and statistics holds, the model may exhibit 
an effective analyticity at low energies. In this case, the conductance is 
completely determined by its behavior under modular transformations.
\end{abstract}
\pacs{}
\narrowtext

\section{Introduction}
\label{sec:intro}

The problem of the nature of the phase transitions between the plateaus of the 
Quantum Hall States is one of the most challenging questions in the
physics of strongly correleted Fermi systems. These phase transitions
separate states whose most important physical feature is their inherent
topological nature. Thus, it has been shown that the fluid ground 
states of Quantum Hall systems have a form of topological
order, instead of conventional long range order~\cite{wen-topolgy}. 
Furthermore~\cite{laughlin,haldane,halperin}, the low-lying excitations are 
fractionally charged quasiparticles with fractional statistics~\cite{leinaas}. 
A direct consequence of the topological order is the universal properties of 
the 
edge states~\cite{wen-edge}. In addition to the values of
the Hall conductance, the plateaus are uniquely defined by the quantum
numbers of their excitations. Hence, the effective theories which
describe the plateau behaviors must be topological field theories. It
has been known for some time that Chern-Simons gauge theories give an
accurate description of the plateaus~\cite{zhk}.

It is then natural to ask if the behavior of the system 
at the phase transtition between a pair of plateaus depends on the
properties of the plateaus.
A theory of the transition between Hall plateaus was proposed recently by Kivelson, 
Lee and Zhang (KLZ)~\cite{phasediagram}. This theory explains the rich phase diagram 
seen experimentally. KLZ used a bosonic description of the FQHE and, in this 
picture, the transition between plateaus becomes strongly reminiscent to the 
superfluid-insulator phase transition in a disordered system. The transition beteween 
plateaus appears to be generally a first order transition in the pure system but it 
is smoothed to a second order transtion by effects of disorder. In this picture, the 
stable FQHE plateaus are the phases in which the Hall conductivity 
$\sigma_{xy}=\nu {\frac{e^2}{h}}$ has precisely defined values. 
Here $\nu$ is a fraction whose denominator is always  {\it odd}.
In this state  the longitudinal conductivity vanishes exactly, $\sigma_{xx}=0$ . 
Since in such theories the Hall conductivity is the coefficient of the 
effective Chern-Simons term for the external electromagnetic field, the transition 
between plateaus becomes a transition between states with effective actions for the 
electromagnetic field with different Chern-Simons coefficients. At the transition 
$\sigma_{xx} \not= 0$ (and perhaps universal) and the denominator of 
$\sigma_{xy}$ is {\it even}.
In the language of the renormalization group, the plateaus
are represented by stable fixed points described by effective actions which have a
pure Chern-Simons form\cite{wen,frohlichzee}. 
From this viewpoint it is natural to ask if the 
critical properties at the transtion between plateaus depends on the precise values of 
the Chern-Simons coefficients of the fixed points or if there is some sort of
{\it superuniversality} at work. Kivelson, Lee and Zhang~\cite{phasediagram}
have argued that the localization length exponent at the transition
between plateaus in a disordered quantum Hall system should be the same
for all these phase transitions, while  the longitudinal conductivity at
the transition may be universal although different for different
transitions. Hence, KLZ have argued that these transtions are {\it
superuniversal}. Furthermore,
L{\"u}tken and Ross \cite{ross-lutken,lutken} have 
conjectured that the theories for the FQHE posses an effective 
{\it modular invariance}
which seemingly implies such a superuniversality in the sense that 
all the critical points
have the same critical exponents and the critical conductances are
related by modular transformations. Recent Monte Carlo simulations on a 
related system by Manousakis~\cite{manousakis} seem to support the notion of 
superuniversality.

Recent experimental work by Shahar {\it et.~al.\/}\cite{shahar} on the
transition between quantum Hall and insulating states have revealed an
effective self-duality relating the properties of the system on either
side of the phase transition. This self-duality is observed in a suprisingly broad  
neighborhood of the critical point. This observation gives strong
motivation to study systems with particle-vortex duality. Of course such
a symmetry can emerge near the phase transition even if the microscopic
theory is not self-dual. 

Chern-Simons gauge field theories thus provide a natural description of the 
plateaus of the fractional quantum Hall effect (FQHE) as well as of other
problems in condensed matter physics and field theories in two 
space dimensions. Chern-Simons theories appear in the effective 
theories for the low energy physics~\cite{zhk,read,wz} (in the Landau-Ginzburg 
sense). Their presence makes manifest the breaking of time reversal invariance in 
the underlying system. Chern-Simons theories are also the natural framework for 
statistical transmutation~\cite{wilczek,wu-zee,witten,polyakov,book}. 
In particular, in the FQHE Chern-Simon theories appear in the form of 
exact mappings between theories in which the particles which 
bear the same electric charge but, in general, have different 
statistics~\cite{zhk,lopez,dreamteam}. 
This latter approach has led to two mutually complementary descriptions of the FQHE
and its hierarchies. One is based on a bosonic picture within which a superfluid
analogy plays a central role. The other uses a fermionic picture with an integer Hall
effect as the reference analogy~\cite{jain}. 

Motivated by these considerations, in this paper we consider the critical properties
of phase  transitions in theories of matter coupled to a Chern-Simons gauge field.
Our main goal is to investigate if the critical properties of such systems
are altered in any relevant way by the coupling of the order 
parameter field to a fluctuating Chern-Simons gauge field. In contrast with the
generally accepted assumption that the transition between plateaus is necessarily
a first order transition in a pure system, we
construct a model in which the transition is apparently 
second order, even in the abscence of disorder. This model can serve as the prototype 
of the phase transition between plateaus. The strategy that we follow in this work 
is to construct physically reasonable models which manifestly have these symmetries 
and to use them to describe the plateaus and the phase transitions. The models should 
also yield the correct electromagnetic response functions at its stable fixed points 
which will then be identified with the plateaus.

More specifically, we consider systems of interacting particles with
fractional statistics parametrized by an angle $\delta$. In Chern-Simons language, this
corresponds to a coupling constant $\theta=1/2\delta$.
In order to address the issue of superuniversality it is necessary to define carefully 
the symmetries exhibited by these systems, even at the microscopic level. 
The problem discussed here is the same as to ask if the universality classes of a phase 
transition depends on the fractional statistics~\cite{spin}. The notion of fractional
statistics is clearly well defined for systems with a finite number of particles
(first quantization). In this case, it is straightforward to check that all 
theories of this type are invariant under a shift of the statistical angle $\delta \to
\delta +2 \pi k$ (where $k$ is an integer) or, equivalently, ${\frac{1}{\theta}} \to 
{\frac{1}{\theta}}+4 \pi k$. This is the invariance of the theory of
the FQHE under the attachment of an even number $2k$ of flux quanta to each particle. 
We will call this symmetry {\it Periodicity} and denote it by $P$ (not to be confused
with parity!).
Also, in systems in which particle-hole symmetry is exact, the theory is invariant 
under a change in the sign of the Chern-Simons coupling constant, $\theta \to -\theta$.
For obvious reasons, we will call this symmetry {\it Charge Conjugation} and denote it
by $C$. Finally, we will show that there is another symmetry at work in these
systems, {\it Duality}, which we will call $S$. However, as in other situations in
statistical physics, {\it Duality} is not a symmetry but rather it maps different
theories among each other. We show however that it is possible to construct
exactly self-dual theories which will serve as fixed points.

In this paper we construct two classes of models, which we call Model I and II 
respectively. Both models exhibit quantum critical behavior at zero temperature,
In Model I, at least, this behavior is associated with zero temperature
phase transitions. Model II appears to be critical at zero temperature for all 
physically meaningful values of the coupling constants.
Both types of model are explicitly Periodic in the statistical angle and, 
in addition, Model II is explicitly self-dual.

Model I describes a system of charged particles with (in
general) fractional statistics and short range interactions. These
models are not self-dual in the sense that the form of the interaction
is not in general preserved by the Duality transformation. In fact,
there are cases in which a theory of particles with short range
interactions is dual to a theory of vortices (the particles of the dual
theory) which have long range interactions. This is in fact the familiar
case of a superfluid-insulator transition. In general, the stable phases of the
systems described by Model I have the qualitative properies of Quantum
Hall plateaus and it is qualitatively similar to the Landau-Ginzburg theory of the
FQHE. This model gives the correct plateau behavior. However, in general,
it is not self-dual and its properties at the phase transtion betweeen
plateaus are not understood. Recently, Pryadko and Zhang~\cite{pryadko} and 
Balachandran, Chandar and Sathialpan~\cite{bala} have
considered a model closely related to Model I and discussed its properties 
under duality transformations. 
Our conclusions on the physics of these model agrees with theirs. 

The physics of the transitions between Quantum Hall plateaus
strongly suggests that, at least very close to the transition, the
system should be effectively self-dual. This motivates the construction
of a model, Model II, which is explicitly self-dual. However, Model II
will turn out to have interactions which are long ranged 
and retarded. Physically, Model II describes the behavior of a system
whose excitations are charged particles with fractional statistics,
restricted to move on a plane. The long-range retarded interactions are
due to the quantum fluactuations of the electromagnetic field in the
full $3+1$-dimensional space-time~\cite{electro}. Naturally, when 
the coupling constant of the long-range interactions becomes very small, 
the short range interactions (of the sort described by Model I) become dominant.

We will show that Model II exhibits quantum critical behavior of a sort that is 
readily analyzed.
We conjecture that this behavior is characteristic of the behavior of
a broader class of systems {\it at their zero temperature
critical points}, including 
Model I.  Model II is manifestly periodic under
shifts of the statistical angle $\delta$. We find it convenient to define a parameter
$f=\delta/2 \pi$. The interactions between these
particles is represented by a long-ranged kernel which at long
wavelenths we choose to be of the form $2\pi g/\sqrt{k^2}$, where $k_\mu$
is the three-momentum ({\it i.~e.\/}, momentum and frequency). The
parameter $g$ is be the coupling constant and, for stability, it must
be positive (repulsive). This model is both self-dual and periodic. We show that
the partition function of Model II is invariant under the {\it Duality} transformations
\begin{equation}
g \to {\frac {g}{g^2+f^2}}  \qquad  \qquad f \to -{\frac {f}{g^2+f^2}} 
\label{eq:dual}
\end{equation}
as well as under {\it Periodicity}
\begin{equation}
f \to f+1
\label{eq:perio}
\end{equation}
These two transformations do not commute with each other. 
Instead, they generate an infinite  discrete group known as the {\it Modular Group} 
or $SL(2,Z)$. Thus these systems are invariant under that action of $SL(2,Z)$.
In fact, in terms of the complex variable $z=f+ig$ (which has ${\rm Im}
z=g >0$), the group of Modular transformations becomes
the group of fractional linear transformations
of the complex plane with integer coefficients
\begin{equation}
z \to {\frac{az+b}{cz+d}}
\label{eq:mod}
\end{equation}
where $a$, $b$, $c$ and $d$ are integers such that $ad-bc=1$.

Since Model II is manifestly invariant under both Periodicity and
Duality, its partition function is  {\it Modular Invariant}. The Modular Group
$SL(2,Z)$ has a rich structure. In particular it has an infinite set of
fixed points, namely points of the upper half plane which are invariant
under at least one, non-trivial modular transformation. We will refer to these points as the 
{\it modular fixed points}. 
L{\"u}tken and Ross\cite{lutken} have used this structure 
as the basis for a conjecture concerning the behavior of possible {\it
renormalization group flows} in QHE systems. However, in the work of L{\"u}tken 
and Ross, $SL(2,Z)$ acts directly on the complex conductance
$\sigma_{xy}+i\sigma_{xx}$. Consequently, the fixed points of $SL(2,Z)$
in reference \cite{lutken} {\it are} the allowed values of the complex
conductances. In contrast, in Model II the modular fixed points are
points of special symmetry. In analogy with the Kramers-Wannier argument for the 
two-dimensional Ising Model\cite{duality}, we can expect the phase transitions
to occur at the modular fixed points. Nevertheless, we will show below that 
{\it Modular Invariance} is enough to determine the conductance of Model
II, at least at the modular fixed points. 
However, we will also find that
there are other values of $f$ and $g$ where the system is critical.

Having an explicit construction of a model which is  
modular invariant has allowed us to investigate the consequences of
modular invariance on physical observables. In particular, we show that
the current correlation functions of these models transform (almost)
like a {\it modular form}. More precisely, we show that from the current-current
correlation function of this system it is possible to construct an amplitude,
 which we call $D(z)$, which determines the complex conductance. By construction,
$D(z)$ is Periodic under $z \to z+1$ and, 
under Duality, it transforms like $D(-1/z)=z^2 D(z)+z$. Combining both
transformations we find that, under a general modular transformation, $D(z)$
transforms like 
\begin{equation}
D({\frac{az+b}{cz+d}})=(cz+d)^2 D(z)+c(cz+d)
\label{eq:modf}
\end{equation}
This functional relation is a direct consequence of the modular
invariance of the partition function and of the fact that 
this invarience is broken by an external
electromagnetic field in a well defined manner. In fact, although we
derived this functional relation in the context of Model II, we expect it
to be generally valid for any modular invariant system. What is very
interesting is the fact that
this functional relation is sufficient to determine the values
of the conductances for all the fixed points of the modular group
$SL(2,Z)$. Below, we derive  a formula which yields the allowed values of the 
conductance at the modular fixed points. Furthermore, by making
additional assumtions on the possible analytic properties of the
correlation functions compatible with the symmetries of the system, we
can also determine their behavior away from the modular fixed points. 
In particular, modular invariance requires that, quite
generally, $\sigma_{xx}$ is finite, {\it at least} at the modular fixed
points. 

Interacting charged particles without impurities are
generally expected to be either insulators or {\it perfect} metals (or even
superconductors). Correspondingly $\sigma_{xx}=0 $ for an insulator
while $\sigma_{xx}=\infty $ for a perfect metal (or
superconductor). In order to find $\sigma_{xx}\not= 0 $ (at zero temperature)
it is usually assumed that the system must be disordered (or break
galilean invariance in some way). However, there is an alternative
scenario which yields a finite $\sigma_{xx}$ in a non-disordered system:
a Quantum Critical Point. At a Quantum Critical Point, the spectral functions of all 
the correlation functions are changed drastically by the effects of
fluctuations on all length scales. In particular, the pole structure associated
with the existence of well defined elementary excitations is wiped out and replaced by 
branch cuts. Hence, the spectrum becomes a broad incoherent continuum. A
direct consequence of this behavior is that $\sigma_{xx} $ can now be
finite. Thus, we will interpret having a finite $\sigma_{xx} $ in a (pure) system of
charged interacting particled as a {\it signature} of a Quantum Critical Point.

Some time ago C.~J.~Callan and D.~E.~Freed found in their work on the 
Dissipative Hofstadter Model\cite{denise} an algebraic structure very similar 
to the one we use in this paper.
It is clear that Model II is quite analogous to the Dissipative Hofstadter 
Model. In particular, our coupling constant $g$ plays the role of the
dissipation in the Dissipative Hofstadter Model. Similarly, the Modular
Group also played a central role in the work of Shapere and Wilczek
\cite{shapere} and of Rey and Zee\cite{rey} on self-dual $Z_N$ models in
$2+1$ dimensions.

The paper is organized as follows. In Section~\ref{sec:models} we
introduce a class of  models which are manifestly invariant under periodic shifts of
the statistical angle. In Section~\ref{sec:duality} we construct the
Duality transformation. In particular, we discuss the way the presence of external
electromagnetic perurbations alter the way these systems transform under Duality.
In Section~\ref{sec:resp} we discuss the electromagnetic response
functions and their properties under Duality. In Section~\ref{sec:phase}
we introduce Model I and Model II and discuss the nature of their phase
diagrams. This section contains the most important results of this
paper, including the determination of the complex conductance at the
modular fixed points. In Section~\ref{sec:QFT} we discuss the role of
periodicity in quantum field theories of matter coupled to Chern-Simons.
Section~~\ref{sec:conc} is devoted to the conclusions.

\section{Models}
\label{sec:models}

For all the reasons invoked above we will adopt a lattice formulation in which all
symmetries are manifest. We will work in discretized $2+1$ dimensional
Euclidean space-time. We consider a three-dimensional cubic lattice with sites $\{
{\vec r} \}$ and links $\{ {\vec r}, \mu\}$ ($\mu=1,2,3$ where $1$ and
$2$ are the space directions and $3$ is (imaginary) time). We represent the worldlines
of the charged particles by means of an integer valued conserved vector field 
$\ell_\mu({\vec r})$ defined on the links of this lattice. 
Conservation means,
naturally, that $\Delta_\mu \ell_\mu({\vec r})=0$ at every site ${\vec r}$, where 
$\Delta_\mu \ell_\mu$ is the lattice divergence. 
The current loops can be
either the worldlines of first quantized particles relative to the empty state or
the worldlines of excitations relative to a reference ground state.
Here we envision several physically
distinct situations. One system we want to describe  cosnsits of non-relativistic 
charged particles with some statistics  and at finite density. In this case, the
worldlines do not end and only close around the time dimension. Another set of cases of
interest are relativistic systems (at zero density). In these cases, closed loops are
allowed and represent particle-antiparticle pairs. Relativistic field
theories are Charge Conjugation invariant. Non-relativistic systems with or without
Charge Conjugation symmetry fit this description as well if one now thinks of the
loops as representing quasiparticle-quasihole pairs. When we compute the
electromagnetic response of these systems, we will have to bear in mind that,
in general, there is a non-trivial contribution from the reference vacuum 
which must be added to the contributions of the excitations.

Thus, we are interested in the behavior of theories which have both closed loops and 
with loops that close around the time dimension. The loops can be described in
terms of a set of {\it bare} particles which can be taken to 
be {\it bosons} but their {\it actual} statistics is determined by a topological 
term in the action. Typically, this is accomplished by coupling matter to a 
Chern-Simons gauge field~\cite{zhk}. Since we want to make all symmetries explicit we
choose to define the theory on a lattice. Chern-Simons theories have been defined on
lattices but they are rather cumbersome to work with. Thus, 
we use an equivalent construction in which we directly attach fluxes to particles and, 
in this way achieve statistical transmutation~\cite{wilczek}. 
The simplest way to incorporate  statistical transmutation in a discretized picture is 
to modify the weight of the path integral by a factor which is a pure phase equal to 
the statistical angle $\delta$ times the {\it linking number} $\nu_L$ of the 
worldlines of the particles. Theories with the property of charged excitations always
being bound to magnetic charge (or flux) are said to exhibit {\it oblique
confinement}~\cite{witten-oblique}. The Landau-Ginzburg theories of the
FQHE~\cite{zhk} are simplest (and best understood) example of oblique 
confinement~\cite{girvin}.

In order to keep track of particles and fluxes, and to avoid ambiguities, 
we {\it define} the worldlines of the {\it fluxes} in such a way
that they strictly follow the matter worldlines but exist on neighboring sites of the 
{\it dual lattice}, which we denote by $\{{\vec R} \}$. Thus, the configuration
of fluxes on the links of the dual lattice is the same as the configuration of matter 
on  the links of the direct lattice, {\it i.~e.\/} $\ell_\mu({\vec r})=\ell_\mu({\vec
R})$ where ${\vec R}\equiv {\vec r} + ({\frac{1}{2}},{\frac{1}{2}},{\frac{1}{2}})$. 
Thus, the vector field $\ell_\mu$ represents {\it both} the world lines of the
particles {\it and} the worldlines of the fluxes which, therefore, do not have an
independent existence. As a consequence, charges cannot detect the constituent
fluxes of other particles without detecting their charge.

Notice that, if the fluxes are attached to the worldlines of the particles by a rigid 
translation, then the system is defined in such a way that the particle cannot feel its own 
flux. In this case there will be no  phase factor associated with the {\it writhing} number 
(or selfwinding)  of the worldline. Hence, 
there is no fractional spin  associated with the configurations but there will be 
fractional statistics. However, one can imagine other definitions of
the system at short distances. For instance, it is possible to construct
the flux-charge composite in such a way that the flux rotates around the
worldline as the latter winds around itself. In this case the  flux-charge composite
becomes the standard rubber band construction associated with the
framing of knots. It is well known from the work of Polyakov~\cite{polyakov} and 
Witten~\cite{witten}
that this construction leads to the conventional spin-statistics
connection. With our explicit construction there is no fractional spin, so
the theory has the invariance $\theta \to - \theta$ discussed in the 
Introduction.
We will show below that the presence of this symmetry imposes severe
constraints on the behavior of the correlation functions.
Conversely, if a fractional spin is associated with the flux-charge
composite, this symmetry will be broken by an amount determined by the
fractional spin.  (We return to this point below.)

The action $S[\ell_\mu]$ for the field $\ell_\mu({\vec r})$ will be chosen to be 
of the form
\begin{eqnarray}
S[\ell_\mu]=&{\frac{1}{2}}& \sum_{{\vec r},{\vec r}'}\; \ell_\mu({\vec r})\; 
G_{\mu \nu}({\vec r}-{\vec r}')\;\ell_\nu({\vec r}')+
{\frac{i}{2}}\sum_{{\vec r},{\vec R}}\; \ell_\mu({\vec r})\; 
K_{\mu \nu}({\vec r},{\vec R})\;\ell_\nu({\vec R})\nonumber \\
&+& i\;  \sum_{{\vec r},{\vec r}'} e({\vec r}-{\vec r}')\;\ell_\mu({\vec r})\; A_\mu({\vec r}')+
 \sum_{{\vec R},{\vec R}'} h({\vec R}-{\vec R}')\; \ell_\mu({\vec R}) B_\mu({\vec R}')\nonumber \\
&+&
{\frac{1}{2}} \sum_{{\vec r},{\vec r}'}\; A_\mu({\vec r})\; 
\Pi_{\mu \nu}^0({\vec r},{\vec r}') \; A_\nu({\vec r}') \ \ .
\label{eq:action1}
\end{eqnarray}
The partition function is
\begin{equation}
{\cal Z}=\sum_{\{\ell_\mu\}} \prod_{\vec r} \delta[\Delta_\mu \ell_\mu(\vec r)] \; 
e^{-S[\ell_\mu]}
\label{eq:pf1}
\end{equation}
where $\{{\vec r}\}$ are the sites of the direct lattice while $\{{\vec R}\}$
 are the sites of the dual lattice, {\it i.~e.\/} 
${\vec R}={\vec r}+({\frac{1}{2}},{\frac{1}{2}},{\frac{1}{2}})$. Here $A_\mu({\vec r})$
is an external electromagnetic field and $B_\mu({\vec R})$ is the corresponding field
strength. $A_\mu({\vec r})$
couples to matter through two coupling constants, an {\it electric} charge $e$ and a
{\it magnetic} charge $h$. 

The physical interpretation of each term of the action of Eq.~(\ref{eq:action1}) 
is as follows:
\begin{enumerate}
\item
The matter interactions are represnted  by 
$G_{\mu \nu}({\vec r}-{\vec r}')$. This is the parity {\it even} part of the action
and it describes the repulsive forces among the particles. For instance, the short
range part of $G_{\mu \nu}({\vec r}-{\vec r}')$ essentially  suppresses configurations
with $|\ell_\mu|>1$ which means that it penalizes the occupancy of one link
by more than one particle. Physically, $G_{\mu \nu}({\vec r}-{\vec r}')$
is the propagator of some field that mediates the interactions between
the particles.
The explicit form of the kernel $G_{\mu \nu}({\vec r}-{\vec r}')$  will be given below. 
\item
The statistical interactions induced by 
Chern-Simons gauge  fields are represented by the tensor 
$K_{\mu \nu}({\vec r},{\vec R}')$ and it is parity {\it odd}. 
The precise form on a lattice of this tensor is rather complicated (see for instance 
reference~\cite{dst}). For our purposes, it will be sufficient to require that
it assigns the correct phase factors depending on the braiding of the worldlines.
The continuum Chern-Simons action (for a gauge field ${\cal A}_\mu$) is
$S_{\rm CS}({\cal A}_\mu)={\frac{\theta}{2}}\epsilon_{\mu \nu \lambda}{\cal A}_\mu
\partial_\nu {\cal A}_\lambda$, where $\theta$ is the Chern-Simons coupling constant.
Then~\cite{witten,polyakov} the statistical angle $\delta$ and the coupling
constant $\theta$ are related by
$\delta=1/(2\theta)$. The weight for a configuration is $\exp(i \delta \nu_L)$, where
$\nu_L$ is the linking number of the configuration~\cite{polyakov,book}.
Using a magnetostatic analogy, it is possible to find~\cite{polyakov} a simple formula
which relates  the linking number $\nu_L$ to a set of currents $J_{\mu}$ which describe 
the worldlines. Explicitly one finds~\cite{polyakov}
\begin{equation}
\nu_L=\int d^3x \; \int d^3y \; J_{\mu}(x)  \epsilon_{\mu \nu \lambda}\; G_0(x-y) \;
\partial_\lambda J_{\nu}(y) \ \ ,
\label{eq:braids}
\end{equation}
where $G_0(x-y)$ is the three-dimensional green's function,
\begin{equation}
-\nabla^2 G_0(x-y)=\delta(x-y) \ \ .
\label{eq:gf1}
\end{equation}
Thus we {\it define} $K_{\mu \nu}$ so that, to each
configuration of matter, the weight of the path-integral has a {\it phase factor} with 
argument equal  to the {\it linking number} of the configuration
up to a coupling constant which we identify with the statistical angle $\delta$.
In other words, the term involving the tensor $K_{\mu \nu}$ represents fractional 
statistics. By construction, it is automatically invariant under periodic shifts of the 
statistical angle.
The explicit form of $K_{\mu \nu}$ depends on details of the discretization procedure.
We will use an explicit form which works provided that the arguments of $K_{\mu \nu}$ 
are defined to be one on the  direct lattice and the other on the dual lattice. 
In  continuum description, a point-splitting regularization is needed to define the 
linking number. 
The lattice regularization takes care of this problems automatically.
\item The terms in the second line of the action of eq.~\ref{eq:action1}   
 represent the coupling to an external vector field 
$A_\mu({\vec r})$. The space components, $1$ and $2$, of this vector field can be 
used to implement an  external magnetic field. 
The time component ($3$) is a scalar potential and a constant piece represents a 
chemical potential (up to an analytic continuation) since it counts the net charge
flowing forwards in time  at a given time slice. 
In general, $A_\mu({\vec r})$
will be an external electromagnetic field that we will use to probe the system. The {\it
electric} coupling $e({\vec r})$ is the conventional lattice version 
of the usual current coupling which we will assume to be mildlyly non-local.
The coupling of the {\it vortex current} to the field strength $B_\mu$, $h({\vec R})$, 
is natural as it represents the coupling between the field and and the circulating  current 
associated with  the vorticity (in the sense of Amp{\`e}re's Law). 
In any event, this term gets generated
automatically under duality even if it was not present in the starting hamiltonian.
Similarly, we have allowed  these couplings to be  non-local since they will generally become 
non-local under duality transformations. In any case at the fixed points that we will find the non
locality of the electric coupling is irrelevant.
\item
The last term of the action represents the electromagnetic response of background degrees of
freedom which do not participate of the dynamics but which affect the electrodynamics of the
system ({\it e.~g.\/}, the conductances, etc.)
\end{enumerate}

Since we are interested in configurations of loops (worldlines) which are
large and well separted from each other, we do not need to specify the details
of the construction of the kernels $G_{\mu \nu}$ and $K_{\mu \nu}$ at short distances. 
The precise form of these kernels at distance
scales of the order of the lattice constant will not affect the universal, long
distance, properties. Thus, in practice, we will only specify their behavior 
at long distances or small momenta. The only
assumption we will make here is that the models of interest can be defined on a lattice,
in such a way that all the symmetries are preserved.

\section{Duality}
\label{sec:duality}

The strategy of this paper is thus to classify the stable phases and phase
transitions of theories with the action of eq.~(\ref{eq:action1}) (and  a number of
simple generalizations). The most direct way to do this is to use duality
transformations~\cite{duality}.  Duality has been used with great success in the study 
of theories with a structure similar to the one we discussed here
~\cite{duality2,duality3,duality4}. 
Of particular importance to the problems discussed
here, is the work of Cardy and Rabinovici~\cite{cardy} on theories of two-dimensional 
statistical mechanics with oblique confinement. Cardy and Rabinovici showed that,
in the context of their models, the phase diagram could be derived by means of a
sequence of transformations: duality, charge conjugation and parity. They were
the first to show that, together, these three transformations generated the group
$SL(2,Z)$ and that the phases are characterized by condensates with oblique
confinement. The phase diagram of the FQHE exhibits, phenomenologically, a very similar
structure. Although a microscopic model to motivate such constructions for systems
related to the FQHE has not been available, the analysis of the phase diagram by 
Kivelson, Lee and Zhang~\cite{phasediagram} and by Ross and
L{\"u}tken~\cite{ross-lutken}
has followed a line very close to that of Cardy and Rabinovici~\cite{cardy}.  

We now proceed to derive a duality transformation of this theory by standard
methods~\cite{duality,duality2,duality3,duality4}. This amounts to rewriting the theory 
in terms of a 
dual theory defined in terms of loops on the dual lattice. Since the flux loops 
reside on the
links of the dual lattice, duality, as usual, is the same as to exchange electric 
currents with magnetic fluxes.
By means of the Poisson Summation formula, the partition 
function can be written in the form
\begin{equation}
{\cal Z}=\sum_{\{m_\mu\}}\int {\cal D}\ell_\mu 
\prod_{\vec r} \delta(\Delta_\mu \ell_\mu) \; 
e^{-S[\ell_\mu]+2\pi i \sum_{{\vec r}} m_\mu(\vec r)\ell_\mu(\vec r)}  \ \ ,
\label{eq:pf2}
\end{equation}
where $m_\mu(\vec r)$ takes values on the integers and $\ell_\mu(\vec r)$ are now 
real numbers. 

We will assume that the system obeys {\it periodic} boundary conditions  in the 
time direction ($x_3$). For simplicity, we will also assume that the system obeys periodic
boundary conditions in space (this choice does not affect the physics of this problem in
any significant way).
The configurations of loops can be classified by the
number of times  they wind around the time direction ($x_3$). Since we
will always work in the thermodynamic limit and will only be interested in systems at
zero total center of mass momentum and current, it is not necessary to consider
configurations which wind around the space directions. However, if we want to
describe systems at non-zero density ({\it i.~e.\/} without Charge Conjugation
invariance) we need to consider configurations with non-zero winding number around the 
time direction.

The constraint on the field $\ell_\mu(\vec r)$ can be solved in terms of a
topologically trivial piece and a field $\ell_\mu^{\rm top}(\vec r)$ which accounts 
for the the loops with non-zero winding number. The topological loops are important to
describe systems at non-zero matter density . The constraint
is solved by
\begin{equation}
\ell_\mu(\vec r)=\epsilon_{\mu \nu \sigma} \Delta_\nu \phi_\sigma(\vec R)+
\ell_\mu^{\rm top}(\vec r) \ \ .
\label{eq:solve}
\end{equation}
In what follows, the topological loops $\ell_\mu^{\rm top}$ will be set to zero.
Notice that, in solving the constraint, a gauge invariance has now appeared; 
since if $\phi_\sigma(\vec R)$ solves the 
constraint,  so does $\phi_\sigma(\vec R)+\Delta_\sigma F(\vec R)$, 
where $F(\vec R)$ is an arbitrary function of $\vec R$. This gauge symmetry of the
dual theory will require the use of a gauge fixing condition when we integrated out
the variables $\phi_\sigma$. 

We now carry out the duality transformation. It is convenient to do it in Fourier space.
In terms of the Fourier transforms of $\ell_\mu$ and $\phi_\mu$
(with wavevector $ k_\mu$), we get
\begin{equation}
\ell_\mu(k)=i \epsilon_{\mu \nu \sigma} \Delta_\nu(k)  \phi_\sigma(k) \ \ ,
\label{eq:solve2}
\end{equation}
where
\begin{equation} 
\Delta_\nu (k) \equiv -i (e^{ik_\nu}-1) 
\label{eq:Delta}
\end{equation}
is a lattice derivative. From now on we will use the notation
 $\Delta_\nu(k) \equiv k_\nu$. 

We now specify the kernels $G_{\mu \nu}$ and $K_{\mu \nu}$. 
Without any loss of generality, the Fourier transforms of the kernels, 
$ G_{\mu \lambda}( k)$ and $K_{\mu \lambda}( k)$ can be taken 
to be of the form
\begin{eqnarray}
{\tilde G}_{\mu \lambda}(k)&=&g( k) \; T_{\mu \lambda}(k)
\nonumber \\
{\tilde K}_{\mu \lambda}(k)&=&{\frac{f( k)}{\sqrt{k^2}}} 
\; C_{\mu \lambda}(k) \ \ ,
\label{eq:kernels}
\end{eqnarray}
where $T_{\mu \lambda}(k)$ and $C_{\mu \lambda}(k)$ are the parity-even and parity-odd
(Chern-Simons) tensors,
\begin{equation}
T_{\mu \lambda}(k)=\delta_{\mu \lambda}-{\frac{k_\mu k_\lambda}{k^2}} \qquad 
C_{\mu \lambda}(k)=i \epsilon_{\mu \lambda \rho}
{\frac{k_\rho }{\sqrt{ k^2}}} \ \ .
\label{eq:Cmunu}
\end{equation}
The coupling functions $g(k)$ and $f( k)$ define
the theory. Notice that since $\ell_\mu(k)$ is a conserved current, a term in
$G_{\mu \nu}$ proportional to the $k_\mu k_\nu$ does not actually contribute
to the action.

We will consider two models, defined by two distinct choices
for $g(k)$. 
In the first case, which we will call Model I, we choose $g( k)\equiv g$. 
This choice describes systems with short range interactions.
In the other case, which we call Model II, we choose ${\tilde G}_{\mu \nu}( k)$ to 
be long ranged, {\it i.~e.\/} $\lim_{{ k_\mu} \to 0} {\sqrt{{k}^2}} g({ k})
={g}$. As explained in the Introduction, this choice corresponds to the non-local (retarded) 
$ 1/|x|^2$ interactions induced by the quantum fluctuations of the electromagnetic field (in the 
full $3+1$-dimensional space-time) on the particles that move on the plane. 
We will show below that Model II is actually a self-dual theory.
The coupling constants of these theories will be given by the zero momentum limits of
the functions. 

We now complete the duality transformation. The dual theory
is defined in terms of the integer-valued dual field $L_\mu(\vec R)$ 
\begin{equation}
L_\mu(\vec R)=\epsilon_{\mu \nu \sigma}\Delta_\nu m_\sigma(\vec r) \ \ .
\label{eq:Lmu}
\end{equation}
The dual theory is found by integrating out the field $\phi_\mu$. 
Because this theory is gauge invariant, we must add a gauge 
fixing term in order to obtain a finite result. The final answer is, naturally, 
independent of the choice of gauge~\cite{foot}. 

Explicitly, we find that the partition function can be written in the form
\begin{equation}
{\cal Z}_{\rm D}[A_\mu(\vec r)]= {\cal Z}_0 \sum_{\{L_\mu\}}e^{-S_{\rm D}[L_\mu]} 
\prod_{\vec R} \delta(\Delta_\mu L_\mu(\vec R)) \ \ .
\label{eq:ZD}
\end{equation}
 $S_{\rm D}[L_\mu]$ is the dual action  and it is given by
\begin{eqnarray}
S_{\rm dual}[L_\mu]&=&{\frac{1}{2}} \sum_{{\vec R},{\vec R}'}\; L_\mu({\vec R})\; 
\left(G^{\mu \nu}_D({\vec R}-{\vec R}')+
i K^{\mu \nu}_D({\vec R},{\vec r}')\right)\;L_\nu({\vec R}')
\nonumber \\
&+&
i\sum_{{\vec R},{\vec r}'} e_D({\vec R}-{\vec r}')\; L_\mu({\vec R})\; A_\mu({\vec r}')+
\sum_{{\vec r},{\vec r}'} h_D({\vec R}-{\vec R}')\; L_\mu({\vec R})\; 
B_\mu({\vec R}')
\nonumber \\
&+&
{\frac{1}{2}} \sum_{{\vec r},{\vec r}'}\; A_\mu({\vec r})\; 
\Pi_{\mu \nu}^D({\vec r},{\vec r}') \; A_\nu({\vec r}') \ \ .
\label{eq:action4}
\end{eqnarray}
The dual kernels and charges of eq.~(\ref{eq:action4}) are in general functions of 
distance and their Fourier transforms are functions of the momentum. We will simplify 
our notation by dropping, from now on, any explicit reference to the momentum  
dependence of the dual kernels and charges. 

The dual kernels are given by
\begin{eqnarray}
G_{\rm D}^{\mu \nu}&=&(2\pi)^2 g_{\rm D} \; T_{\mu \nu}
\nonumber \\
K_{\rm D}^{\mu \nu}&=&- (2\pi)^2 {\frac{f_{\rm D}}{\sqrt{k^2}}}
C_{\mu \nu} \ \ ,
\label{eq:dualkernels}
\end{eqnarray}
where the dual coupling functions $g_{\rm D}$ and $f_{\rm D}$ 
are given by
\begin{eqnarray}
g_{\rm D} &=&(2\pi)^2 
{\frac{g}{f^2+ k^2 g^2}} 
\nonumber \\
f_{\rm D} &=& - (2\pi)^2
{\frac{f}{f^2+ k^2 g^2}} \ \ .
\label{eq:duality-transf}
\end{eqnarray}
Eq.~(\ref{eq:duality-transf}) is the transformation law of the coupling
functions under  Duality. Similarly, the dual electric $e_D$ and magnetic
charges $h_D$ are
\begin{eqnarray}
e_D&=& {\frac{1}{2\pi}}\left(e f_D+k^2 h g_D \right)
\nonumber \\
h_D&=& {\frac{1}{2\pi}}\left(-e g_D+h f_D \right) \ \ .
\label{eq:functions}
\end{eqnarray}

$\Pi_{\mu \nu}^D$ is the polarization tensor of the dual theory in regimes
in which the dual loops are suppressed. Thus, it is an effective renormalization of
the tensor $\Pi^0_{\mu \nu}$. In general, the full polarization tensor includes
contributions coming from both the loops and from the background. 
$\Pi_{\mu \nu}^D$ has the form
\begin{equation}
\Pi_{\mu \nu}^D= \Pi^{(S)}_D \; T_{\mu \nu}
-i \Pi^{(A)}_D C_{\mu \nu} \  \ ,
\label{eq:pimunud}
\end{equation}
where $\Pi^{(S)}_D$ and $\Pi^{(A)}_D$ are the amplitudes for the parity-even and 
parity-odd pieces of $\Pi_{\mu \nu}^D$,
\begin{eqnarray}
\Pi^{(S)}_D&=& \Pi^{(S)}_0+  k^2\; (e^2 - h^2 k^2)
{\frac{ g_D}{(2\pi)^2}}  -k^2\; 2 h e {\frac{f_D}{(2\pi)^2}}
\nonumber \\
\Pi^{(A)}_D&=&\Pi^{(A)}_0-{\sqrt{k^2}}(e^2 - h^2 k^2){\frac{ f_D}{(2\pi)^2}} 
-k^2 2 e h   {\frac{ g_D}{(2\pi)^2}}{\sqrt{k^2}}
\label{eq:morepies}
\end{eqnarray}
Hence, the polarization tensor acquires {\it additive} corrections under
a duality transformation. This feature will play a very important role
in the analysis of the conductance.

\section{Electromagnetic Response}
\label{sec:resp}

Using the results of the previous section we can, in principle,  determine the long-distance
properties of the possible fixed points.
We will parametrize the fixed points in terms of their response to electromagnetic 
perturbations,
{\it i.~e.\/} the longitudinal conductivity $\sigma_{xx}$ and the Hall conductivity 
$\sigma_{xy}$ 
for metallic phases and the dielectric constant $\varepsilon$ and the Hall conductivity 
$\sigma_{xy}$ for the insulating states.

As usual, these responses are determined from the current-current correlation functions. However,
the physics of this system is determined naturally by the loop-loop correlation 
functions 
\begin{equation}
{\cal D}_{\mu \nu}({\vec r}-{\vec r}')= 
\langle \ell_\mu ({\vec r}) \ell_\mu ({\vec r}')\rangle \ \ .
\label{eq:loopcorr}
\end{equation}
While ${\cal D}_{\mu \nu}$ is related to the current-current correlation functions, 
it is not
equal to them.
The current correlation functions are instead equal to the full polarization tensor 
for the system
 $\Pi_{\mu \nu}({\vec r}-{\vec r}')$ which is defined by
\begin{equation}
\Pi_{\mu \nu}({\vec r}-{\vec r}')=\langle J_\mu({\vec r})\; J_\nu({\vec r}')\rangle=
- \left. 
{\frac{\delta^2 }{\delta A_\mu({\vec r})\delta A_\nu({\vec r}')}}\;
\log {\cal Z}(\{A_\mu(\vec r)\})\right|_{A_\mu=0} \ \ ,
\label{eq:fullpi}
\end{equation}
where $J_\mu$ is the electromagnetic current.

Since the current loops $\{\ell_\mu\}$ are {\it conserved}, {\it i.~e.\/} $\Delta_\mu
\ell_\mu({\vec r})=0$ at every site $\vec r$, we can always write ${\cal D}_{\mu \nu}$ in terms
of transverse tensors. Thus, in Fourier space we write
\begin{equation}
{\cal D}_{\mu \nu}(k)={\cal D}^{S}(k;g,f) \; T_{\mu \nu}(k)-i
{\cal D}^{A}(k;g,f) \; C_{\mu \nu}(k) \ \ .
\label{eq:calD}
\end{equation}
Here, ${\cal D}^{S}(k;g,f)$ and ${\cal D}^{A}(k;g,f)$ are the parity-even and 
parity-odd amplitudes of ${\cal D}_{\mu \nu}$, $T_{\mu \nu}(k)$ and $C_{\mu \nu}(k)$ are the 
tensors defined in eq.~(\ref{eq:Cmunu}). Once again, it should be kept in mind that 
the kernels and the couplings have an explicit momentum dependence.

The relation between the loop correlation function ${\cal D}_{\mu \nu}(k)$ and the current
correlation function $\Pi_{\mu \nu}(k)$ can be determined by differentiating the partition
function. If we also separate the $\Pi_{\mu \nu}$ into  parity-even and  parity-odd pieces,
\begin{equation}
\Pi_{\mu \nu}(k)=\Pi^{S}(k;g,f) \; T_{\mu \nu}(k)-i \Pi^{A}(k;g,f) \; C_{\mu \nu}(k) \ \ ,
\label{eq:split}
\end{equation}
we find
\begin{eqnarray}
\Pi^{S}(k;g,f)&=& \Pi_0^{S}(k) - (e^2-k^2 h^2) \; {\cal D}^{S}(k;g,f)
- 2 e\; h \; \sqrt{k^2} {\cal D}^{A}(k;g,f)  \nonumber\\
\Pi^{A}(k;g,f)&=& \Pi_0^{A}(k) - (e^2-k^2 h^2) \; {\cal D}^{A}(k;g,f)
+ 2 e\; h\; \sqrt{k^2} {\cal D}^{S}(k;g,f) \ \ .
\label{eq:relation}
\end{eqnarray}
A {\it Parity} transformation is equivalent to the mapping $g \to g$ and $f \to -f$.
Hence, under Parity, the amplitudes ${\cal D}^{S}(k;g,f)$ and ${\cal D}^{A}(k;g,f)$
transform like
\begin{eqnarray}
{\cal D}^{S}(k;g,-f) &=& {\cal D}^{S}(k;g,f) \nonumber\\
{\cal D}^{A}(k;g,-f) &=& -{\cal D}^{A}(k;g,f) \ \ .
\label{eq:amplitudes}
\end{eqnarray}
Thus, the amplitudes $\Pi^{S}(k;g,f)$ and $\Pi^{A}(k;g,f)$ transform in exactly the same
way.

By using the definition of the conductivities as linear response 
coefficients, the components of the conductivity tensor 
$\sigma_{ij}$ ($i,j=1,2$)
 can now be  determined by the standard relations (in imaginary time $x_3$)
\begin{equation}
\sigma_{ij}=\lim_{k_3 \to 0} \ \lim_{{\vec k} \to 0} \ {\frac{1}{k_3}} \Pi_{ij}(k) \ \ .
\label{eq:conductivity-tensor1}
\end{equation}
Hence,
\begin{eqnarray}
\sigma_{xx}&=&\lim_{k_3 \to 0} {\frac {1}{k_3}} \Pi^{S}(0,k_3)
\nonumber \\
\sigma_{xy}&=&\lim_{k_3 \to 0}{\frac {1}{|k_3|}} \; \Pi^{A}(0,k_3) \ \ ,
\label{eq:conductivities}
\end{eqnarray}
where we have set $\sigma_{11}=\sigma_{22}\equiv \sigma_{xx}$ and $\sigma_{12}=-\sigma_{21}\equiv
\sigma_{xy}$. We can now relate $\sigma_{xx}$ and $\sigma_{xy}$ to the coupling constants of this
system
\begin{eqnarray}
\sigma_{xx}&=&\sigma_{xx}^0+  \lim_{k_3 \to 0} {\frac{1}{k_3}} \left[-
(e^2-k_3^2 h^2) \; {\cal D}^{S}(k_3;g,f)- 2 e \; h \; |k_3|
 {\cal D}^{A}(k_3;g,f) \right]
              \nonumber\\
\sigma_{xy}&=&\sigma_{xy}^0+  \lim_{k_3 \to 0} {\frac{1}{k_3}} \left[-
(e^2-k_3^2 h^2) \; {\cal D}^{A}(k_3;g,f)+ 2 e\; h\; |k_3|
{\cal D}^{S}(k_3;g,f) \right] \ \ .
\label{eq:coefficients}
\end{eqnarray}
If the system is in a {\it metallic} phase both conductivities $\sigma_{xx}$ and 
$\sigma_{xy}$ are in general different from zero. If the system has an energy gap and, 
hence, it is incompressible, the longitudinal conductivity $\sigma_{xx}$ vanishes. 
Instead, $\Pi^{S}(k)/k^2$ has a finite limit as $k \to 0$ which can be identified with
 the inverse of the effective dielectric 
constant, $\varepsilon^{-1}\equiv \lim_{k \to 0}\Pi^{S}(k)/k^2$ of the system in an 
{\it insulating} phase. If the insulator has broken Time Reversal symmetry, as in the
FQHE, the Hall conductivity $\sigma_{xy}$ may be non-zero. 

The phase diagram of the system is spanned by the asymptotic behavior at small momenta of the coupling 
functions $g(k)$ and $f(k)$, and of the electric and magnetic charges $e(k)$ and $h(k)$.
In the next section we will discuss two distinct limiting 
behaviors for $g$ which yield quite  different physics. In general, the
conductivity tensor at momentum scale $k$ is a function of $g$, $f$, $e$ and
$h$. In the next section we will exploit the existence of these duality transformations to
determine the properties of both the stable phases of this system  and of the 
phase transitions which separate the phases.

\section{Phase Diagram}
\label{sec:phase} 

In this section we will apply the general duality results of section~\ref{sec:duality}
to Models I and II and  use it to determine the qualitative features of the phase
diagram.
In this paper we will considering only theories with Charge Conjugation invariance. 
In this case the total charge $Q$ is set to zero and only loops that close on 
themselves are taken into account. 

\subsection{Short Range Interactions}
\label{subsec:screened}

Let us consider first the case of Model I. In this model,
the interactions are short ranged  and the effective action is characterized by finite 
limiting values of the coupling constants. Hence,  we choose the
parametrization 
\begin{eqnarray}
g&=&  \lim_{k \to 0}{\frac{ g(k)}{2\pi}}\nonumber \\
f&=&\lim_{k \to 0} {\frac{ f(k)}{2\pi}}
\label{eq:fbar}
\end{eqnarray}
This choice of parametrization reflects the fact that since the interaction is
short ranged, $g(0) {\not=}0$. Also, $f(0)$ is the statistical angle and as
such it is defined modulo $2\pi$.
Thus, {\it Periodicity} is 
the shift of $f$ by integers, {\it i.~e.\/} $f \to f+ n$.

Under {\it Duality}, the coupling constants of Model I transform like
\begin{eqnarray}
{g}_{\rm D} &=& {\frac{{g}}{{f}^2+k^2{g}^2}} \nonumber \\
{f}_{\rm D} &=& -{\frac{f}{{f}^2+k^2{g}^2}}
\label{eq:duality-screened}
\end{eqnarray}
Eq.~(\ref{eq:duality-screened}) shows that {\it if} ${f}\not= 0$, the effective
dual couplings as $k \to 0$ become
\begin{eqnarray}
{g}_{\rm D}&=&{\frac{{g}}{{f}^2}}\nonumber \\
{f}_{\rm D} &=& -{\frac{1}{{f}}}
\label{eq:duality-screened2}
\end{eqnarray}
The electric and magnetic charges obey similar transformation laws
\begin{eqnarray}
e_D&=& {\frac{1}{2\pi}}\lim_{k \to 0}  \left(e f_D(k)+k^2 h g_D(k) \right)=
e f_D
\nonumber \\
h_D&={\frac{1}{2\pi}}&\lim_{k \to 0}  \left(-e g_D(k)+h f_D(k) \right)=
-e g_D+h f_D 
\label{eq:duality-charges}
\end{eqnarray}

Thus the theory is self-dual in the sense that the original theory and its dual have
the same form. Clearly, since by shifts of the stastitical angle it is always possible
to bring ${f}$ to the fundamental period $0<{f}\leq 1$, the effective
dual coupling satisfies the inequality ${g}_{\rm D} \geq {g}$ and the dual
coupling constant is larger that the original one.
One usually expects that, if the original coupling constant is
large, the dual coupling constant should be small;  if this property holds, the
strong and weak coupling regimes are related by duality. In this case, the analysis
of these limits becomes very simple since in one regime the loops are suppressed while
in the other the dual loops are suppressed.
However, in general, the coupling constants of Model I,
defined in eq.~(\ref{eq:duality-screened}) and  ~(\ref{eq:duality-screened2}) ,
do not have this property. Hence, if the coupling constant of the original theory is
of order one, the dual coupling is generally larger than one. Therfore, for Model I
duality does not help (in general) to determine the behavior of the system in
asymptotic regimes.

In addition, if ${f}=0$ (up to integer shifts), the situation is quite different. 
Now the dual coupling is long ranged 
${g}_{\rm D}(k)=1/({ k}^2 {g})$. This behavior leads to logarithmic
interactions of matter loops in real space.
The fact that a theory of bosonic loops with short ranged interactions is dual to a
theory of loops with Biot-Savart-like logarithmic interactions is well known
from the three-dimensional classical $XY$-model~\cite{bkm}. The loops with long range
logarithmic interactions are the vortices of the ordered (or superfluid) phase of the 
$XY$ model. 

\subsection{Self-Dual Actions}
\label{subsec:selfdual}

The transformation properties of Model II under duality can be best studied in terms 
of the dimensionless coupling constants $f$ (the same as in eq.~(\ref{eq:fbar}))
 and $ g$ which is given by
\begin{equation}
\begin{array}{l c l}
f\equiv{\displaystyle{\lim_{k \to 0} {\frac{f(k)}{2\pi}}}}
&\qquad&  
g \equiv {\displaystyle{\lim_{k \to 0} {\frac{{\sqrt{k^2}} g(k) }{2\pi}}}} \ \ .
\end{array}
\label{eq:gbar}
\end{equation}
For space-time separations $x-x'$, the interaction kernel $G_{\mu \nu}(x-x')$
for Model II Model II is transverse and it decays like $g/(x-x')^2$.
This is the propagator for Maxwell photons in the full $3+1$-dimensional
space-time.
Thus, Model II describes a system of charged particles with fractional
statistics moving on a plane interacting with the full quantized electromagnetic field
(see however, the caveats of reference~\cite{electro}).

For Model II, the dual coupling constants $ f_{\rm D}$ and $g_{\rm D}$ are
\begin{equation}
\begin{array}{l c l}
g_{\rm D}={\displaystyle{\frac{g}{f^2+g^2}}} & &
f_{\rm D} = -{\displaystyle{\frac{f}{f^2+g^2}}} \ \ .
\end{array}
\label{eq:duality-II}
\end{equation}
For the case of Model II, the electric and magnetic charges are required to obey
\begin{equation}
\begin{array}{l c l}
e\equiv {\displaystyle{\lim_{k \to 0}}}\;  e(k) 
&\qquad &
h 
\equiv {\displaystyle{\lim_{k \to 0}}} \; {\sqrt{k^2}} h(k) \ \ .
\end{array}
\label{eq:magII}
\end{equation}
The dual electric and magnetic charges $e_D$  and $h_D$ are
\begin{eqnarray}
e_D&=&  e f_D+ h g_D 
\nonumber \\
{h}_D&=& -e g_D+ h f_D \ \ .
\label{eq:chargesII}
\end{eqnarray}
It is convenient to define the complex variable $z\equiv {f}+i{g}$. Clearly, while $f$
can take any value (although only rational values are physically meanigful),
$g$ has to be a positive real number for the interactions to be repulsive and
the system to be stable. Hence, the complex variable $z$ takes values on $H$,
the upper half of the complex plane.

In terms
of $z$, {\it Duality} is the mapping
\begin{equation}
z \to z_{\rm D}=-{\frac{1}{z}} \ \ ,
\label{eq:mapping1}
\end{equation}
In this notation, {\it Periodicity} becomes the mapping
\begin{equation}
z \to z_{\rm D}=z+n \ \ ,
\label{eq:mapping2}
\end{equation}
and {\it Charge Conjugation} is now
\begin{equation}
z \to z_{\rm D}=-z^* \ \ .
\label{eq:mapping3}
\end{equation}
Notice that it is always possible to use Periodicity to make 
${\rm Re}\; z={f}\geq 0$.
Thus, Charge Conjugation is actually redundant.

Hence, Model II is {\it self-dual} in the sense that under Duality it maps 
into a model with interactions of the same form. Moreover, it has the extended 
invariance parametrized by the mappings of 
eq.~(\ref{eq:mapping1}), ~(\ref{eq:mapping2}) and ~(\ref{eq:mapping3}). 
These mappings generate the Modular Group $SL(2,Z)$, the group of
fractional linear transformations ${\cal T}$ with integer coefficients of 
the upper half of the complex plane ${\cal H}$ into itself of the form
\begin{equation}
{\cal T} z={\frac{az+b}{cz+d}}  
\label{eq:modular}
\end {equation}
with $ a,b,c,d \in Z$ and $ad-bc=1$. Clearly, the partition function of 
Model II is Modular Invariant and hence its phase diagram is generated by 
$SL(2,Z)$. 
Thus, we expect that Model II should have the conjectured structure of the phase 
diagram of the  FQHE~\cite{phasediagram,ross-lutken}. We will show below that
Modular Invariance has profound consequences. In particular , it determines the
possible behaviors of physical observables such as the conductances.

We also give the transformation rules of the {\it background} conductivities under
duality transformations. They are
\begin{eqnarray}
\sigma^D_{xx}&=&\sigma^0_{xx}+{\frac{1}{2\pi}}\left[(e^2-h^2)g_D
- 2 e h f_D \right]\nonumber\\
\sigma^D_{xy}&=&\sigma^0_{xy}+{\frac{1}{2\pi}}\left[-(e^2-h^2){f}_D
- 2 e h g_D \right] \ \ .
\label{eq:dual-back}
\end{eqnarray}
It is convenient to rescale the amplitudes of 
Eq.~(\ref{eq:coefficients}) and to define
\begin{equation}
\begin{array}{l c l}
{\displaystyle{d^S({g},{f})= 
{\frac{2\pi}{\sqrt{k^2}}}{\cal D}^S({g},{f})
}}&\qquad& 
{\displaystyle{d^A({g},{f})=
{\frac{2\pi}{\sqrt{k^2}}}{\cal D}^A({g},{f})
}} \ \ .
\end{array}
\label{eq:effamps}
\end{equation}
The self duality of the theory
imposes constraints on rescaled amplitudes $d^S({g},{f})$ and 
$d^A({g},{f})$ and, through them, on the  physical conductivities. 
Thus, from
Eq.~(\ref{eq:coefficients}), we get that the physical conductivities $\sigma_{xx}$
and $\sigma_{xy}$ at coupling constants ${g}$ and ${f}$, for specified electric
and magnetic charges $e$ and $h$ are
\begin{eqnarray}
\sigma_{xx}&=&
\sigma^0_{xx}+{\frac{1}{2\pi}}\left[-({e}^2-{h}^2)d^S({g},{f})
- 2{e} {h} d^A({g},{f})
\right]\nonumber\\
\sigma_{xy}&=&
\sigma^0_{xy}+{\frac{1}{2\pi}}\left[-({e}^2-{h}^2)d^A({g},{f})
+ 2{e} {h} d^S({g},{f})
\right] \ \ .
\label{eq:cond1}
\end{eqnarray}
Likewise, for a Time Reversal Invariant system, we have the further restriction
imposed by
Eq.~(\ref{eq:amplitudes}) on the amplitudes ${\cal D}^S$ and ${\cal D}^A$,
\begin{eqnarray}
d^S({g},-{f})&=&d^S({g},{f})\nonumber\\
d^A({g},-{f})&=&-d^A({g},{f}) \ \ .
\label{eq:symm}
\end{eqnarray}
We stress here that Time Reversal follows from choosing a specific definition
of the loops at short distances in such a way that no {\it fractional
spin} is assigned to a system with a given fractional statistics. 
For systems {\it with fractional spin}, Time Reversal is broken and the 
Reflection Symmetry of Eq.~(\ref{eq:symm}) does not hold.

The transformation properties of the quantities of physical interest become 
very transparent by using complex notation. Thus we define the complex 
conductivity $\Sigma$, the amplitude $D$ and the charge $H$,
\begin{eqnarray}
\Sigma(z) &=& \sigma_{xy}(f,g)+i \sigma_{xx}(f,g)\nonumber\\
D(z) &=& d_A(f,g)-id_s(f,g)\nonumber\\
H &=& h-ie \ \ ,
\label{eq:complex}
\end{eqnarray}
where, for simplicity, we have used the notation $D(z)=D(z,z^*)$.
( Here the {\it stars} denote complex conjugation.)  Naturally, this notation 
does not imply that $D(z)$ is necessarily only a function of $z$ and thus that 
it is analytic. In general, the amplitude $D$ is a separate function of $z$ 
and $z^*$  since, {\it a priori}, there is no reason to expect $D$ to be 
an analytic function of the coupling constants $f$ and $g$.
However, as we will see below, there are situations of physical interest where
it can be locally analytic. We will also see that analyticity
has far reaching consequences. In what follows we will write all functions of 
$f$ and $g$ as functions of $z$. (The same caveats on analyticity still apply.)

With this notation, the complex conductivity $\Sigma$ becomes
\begin{equation}
\Sigma(z)=\Sigma_0+ {\frac{H^2}{2\pi}} D^*(z) \ \ .
\label{eq:Sigma}
\end{equation}
Similarly, we find that the duality transformation for the charges is now simply 
\begin{equation}
H_D=H \; z^*_D \ \ .
\label{eq:dualH}
\end{equation}
Since this is a self-dual thery, the conductivity of the theory at $z$ and at $z_D$ 
{\it must be the same}. Hence
\begin{equation}
\Sigma(z)=\Sigma_D+ {\frac{H_D^2}{2\pi}} D^*(z_D) \ \ ,
\label{eq:dualSigma}
\end{equation}
where
\begin{equation}
\Sigma_D=\Sigma_0+{\frac{H^2}{2\pi}} z_D^* \ \ .
\label{eq:SigmaD}
\end{equation}
The consequence of these relations is the the amplitude $D(z,z^*)$ obeys the 
following  functional equation
\begin{equation}
z^2D(z)+z=D(-{\frac{1}{z}}) \ \ ,
\label{eq:functional}
\end{equation}
which is the transformation law of the amplitude $D(z)$ under duality. 
In addition, $D(z)$ must {\it periodic} which here means
\begin{equation}
D(z)=D(z+1) \ \ .
\label{eq:periodicity}
\end{equation}

Similarly, the {\it Charge Conjugation} symmetry transformation of the 
amplitude under  $(f,g) \to (-f,g)$, Eq.~(\ref{eq:symm}), which only holds if 
the system is invariant under {\it Time Reversal}, becomes now the 
{\it Reflection} symmetry,
\begin{equation}
D(-z^*)=-D^*(z) \ \ ,
\label{eq:mirror}
\end{equation}
which expresses the invariance under Time Reversal. Clearly, $D(z)$ {\it
cannot} be analytic if the system is Time Reversal Invariant.

In the next subsection we will use Modular Invariance (and Time Reversal
when appropriate) to determine constraints on the allowed values of
$D(z)$. Once $D(z)$ is known, the conductance $\Sigma(z)$ is determined
uniquely by Eq.~(\ref{eq:SigmaD}) once a choice of the electromagnetic
coupling $H=h-ie$ is made. In what follows, we will choose the {\it
bare} model to have $H=-ie$ ({\it i.~e.} $h=0$). 

\subsection{Consequences of Modular Invariance}
\label{subsec:modular}

Let us now discuss the consequences of {\it Modular Invariance}. 
The Modular Group $SL(2,Z)$ is an infinite, discrete, non-abelian group. 
It has two generators, which we will take to be the Periodicity $P$ and the 
Duality $S$.  Please notice that in the mathematics
literature\cite{modular}, as well as in String
Theory\cite{stringduality}, the convention is to call $T$-duality what we call
Periodicity  and $S$-duality what we call Duality.
By the action of $SL(2,Z)$, it is always possible to map an arbitrary 
point of the upper half-plane ${\cal H}$ into the Fundamental Domain 
$F=\{z\in {\cal H} | |z|\geq 1, |{\rm Re}z|\leq {\frac{1}{2}}\}$.
Particular cases include  the mapping the interior of the unit circle $|z|<1$ 
to the exterior $|z|>1$, which is the conventional mapping of weak and strong 
coupling regimes into each other. 

The unit circle  ${\cal C}_1= \{z;|z|=1\}$ plays a special role since it is a 
self-dual manifold although it is not modular invariant.
On the unit circle $f^2+g^2=1$, duality 
reduces to the mapping $f_D=-f$ and $g_D=g$. In particular, the self-dual 
{\it boson} point on the unit circle is $(f,g)=(0,1)\equiv i$ 
and the self-dual {\it fermion} point on the unit circle is 
$(f,g)=(1/2,\protect {\sqrt{3}}/2)\equiv \rho$. The boson and fermion
self-dual points have very special properties. Let $[i]$ and $[\rho]$ denote the 
sets of points of the upper half-plane ${\cal H}$ which are the images
of $i$ and $\rho$ under $SL(2,Z)$ respectively. For similar reasons, we
will call the sets of points with {\it integer} real parts (
{\it i.~e.\/} $f \in Z$) the {\it boson lines}. Likewise, the the sets of points with 
{\it half-integer} real parts {\it i.~e.\/} $f \in Z+{\frac{1}{2}}$) are the 
{\it fermion lines}. In particular, ${\rm Re} [z]=1$ will be {\it the}
boson line while ${\rm Re} [z]={\frac{1}{2}}$is {\it the} fermion
line.

In addition of the finite modular fixed points $[i]$ and $[\rho]$ the
modular group also has fixed points at extreme regimes of the upper half
plane. The extreme fixed points are the point $\{i \infty \}$ and
$[\infty]$,the set $Q$ of rational points on the real axis. The point $\{i \infty \}$ 
is the limit ${\rm Im} z \to \infty$ and corresponds to the strong coupling limit 
$g \to \infty$.  In this regime the loops are suppressed. Conversely, $[\infty]$ is 
the set of points of $\cal H$ with $g=0$ and  $f$ rational. These points map into each 
other under $SL(2,Z)$ and are also images of the point at infinity on the real axis.
Also, since $g=0$, physically the set of points $[\infty]$ corresponds to  systems of 
non-interacting charged particles. Also, since $g \to 0$, for these points it is not
legitimate to neglect other, shorter range, interactions. In fact, close
enough to these points,  Model II behaves like Model I. Clearly, the
behavior near these points is quite simple since, for large values of
$g$ the loops are suppressed while for small values of $g$ the dual
loops are suppressed. 

The bosonic point $i$ and the fermionic point $\rho$ are examples of 
{\it fixed points} of $SL(2,Z)$ and describe interacting systems. 
It can be shown~\cite{rankin} that the set ${\cal C}$ of all 
points of the upper half-plane ${\cal H}$ which are fixed points under the 
Modular Group $SL(2,Z)$ (namely the set of points in ${\cal H}$ which 
are fixed points of {\it a} modular transformation), is the union of $[i]$, 
$[\rho]$ and $[\infty]$,{ \it i.~e.\/} 
$ {\cal C}= [i]\cup [\rho]  \cup [\infty]$. In what follows we will say
that a fixed point is {\it bosonic} or {\it fermionic} if
it is in $[i]$ or $[\rho]$ respectively. Notice that there will be
points on fermion lines which are bosonic since they are images of $i$, 
for example $z={\frac{1}{2}}+{\frac{i}{2}}$.

These three families of {\it modular} fixed points should
determine the qualitative global features of the phase diagram. This
fact has been stressed by Ross and L{\"u}tken~\cite{ross-lutken}, and by 
L{\"u}tken~\cite{lutken} who have given a detailed description of these
modular fixed points. In that work, they proposed a set of 
phenomenological  {\it Renormalization Group} flows and fixed points
which are {\it compatible} with the requirement of modular invariance.
Thus, Ross and L{\"u}tken were led to the conclusion that the RG fixed points 
should also be $SL(2,Z)$ or modular fixed points. 
Thus, the set $[\infty]$ describes {\it
plateaus} while the finite fixed points $[i]$ and $[\rho]$ describe phase 
transitions. The qualitative phase diagram of Model II has the same
qualitative structure as the phase diagram of  L{\"u}tken~\cite{lutken}
(see in particular L{\"u}tken's figure 2). This is also the phase
diagram that has been proposed by KLZ for the Fractional Quantm
Hall Effect~\cite{phasediagram}.
However, we will show evidence that suggests that Model II has 
properties that we would normally assign to critical points
at most points of the parameter space ${\cal H}$.

Before proceeding further it is convenient write down the
transformation law of the complex conductance $D(z)$ for an arbitary
modular transformation. The transformation laws for $D(z)$ under {\it
duality} Eq.~(\ref{eq:functional}) and {\it periodicity} 
Eq.~(\ref{eq:periodicity}),  generalize for an {\it arbitrary} modular 
transformation ${\cal T} \in SL(2,Z)$ of the  form of
Eq.~(\ref{eq:modular}),  to the following law
\begin{equation}
D({\frac{az+b}{cz+d}})=(cz+d)^2 D(z)+c(cz+d) \ \ .
\label{eq:G2}
\end{equation}
Once again, we stress that so far we have not made any assumptions on
the analytic properties of $D(z)$. 

Equation ~(\ref{eq:G2}) has a very striking and suggestive form. 
Since the partition function is invariant under the action of $SL(2,Z)$, 
naively we would have expected that the complex amplitude $D(z)$ should also
 be invariant under $SL(2,Z)$. However, the invariance of the partition
function does not require the invariance of correlation functions.
Instead, only {\it covariance} is actually required. This is precisely
what happens in critical phenomana where scale invariance only requires
that correlation functions to exhibit scaling behavior. Thus,
in principle, we could have anticipated that if $D(z)$ was not a modular
invariant, it would instead transform {\it homogeneously} under the action
of the Modular Group $SL(2,Z)$. Functions $F_{k,{\bar k}}(z,z^*)$, labelled by 
two integers $k$ and ${\bar k}$, that transform homogeneously under $SL(2,Z)$
as
\begin{equation}
F_{k,{\bar k}}({\cal T} z,({\cal T}z)^*)=(cz+d)^k (c^*z^*+d^*)^{{\bar k}} 
F_{k,{\bar k}}(z,z^*) 
\label{eq:form}
\end{equation}
are said to be modular of weight $(k,{\bar k})$. In the special case when the 
function $F(z)$ is holomorphic (and $k={\bar k}$), it transforms like
\begin{equation}
F_{2k}({\cal T} z)=(cz+d)^{2k} F_{2k}(z)
\label{eq:modularform}
\end{equation}
and it is called a {\it modular form} of weight $2k$. Typical examples
of modular forms are the Eisenstein series $G_{2k}(z)$
\begin{equation}
G_{2k}(z)={\sum_{n,m\in Z}}'{\frac{1}{(nz+m)^{2k}}}
\label{eq:eisenstein}
\end{equation}
and their generalizations\cite{rankin}. 

However, Eq.~(\ref{eq:G2}) transforms {\it inhomogeneously}, {\it i.~e.\/} 
in addition to the scale factor of the form of Eq.~(\ref{eq:modularform}) (with
weight $2k=2$), there is an extra additive contribution. This feature can
be traced back to the duality transformation properties of the
polarization tensor. Indeed, as Eq.~(\ref{eq:morepies}) shows, an external 
electromagnetic perturbation (no matter how weak) breaks self-duality. We will 
see shortly that $D(z)$ is related to a (loosely defined) modular form
of weight 2. This inhomogeneous transformation law has a striking
similarity with the transformation law of the energy-momentum (or
stress-energy) tensor in two-dimensional Conformal Field Theory which
exhibits a {\it Conformal Anomaly}\cite{bpz}. By analogy we can regard the
inhomogeneous transformation law of Eq.~(\ref{eq:G2}) as a {\it Modular
Anomaly}.

It is straightforward to show that if $z_0$ is a {\it fixed point}, 
the generalized transformation law 
Eq.~(\ref{eq:G2}) determines uniquely the value of $D(z_0)$ at the
fixed points. Indeed, let $z_0 \in {\cal C}$ and let ${\cal T}_{z_0}$ 
be the modular 
transformation that leaves $z_0$ invariant, {\it i.~e.\/} 
${\cal T}_{z_0} z_0=z_0$. Then Eq.~(\ref{eq:G2}) reads
\begin{equation}
D({\cal T}_{z_0} z_0)=D(z_0)=(cz_0+d)^2 D(z_0)+c(cz_0+d) \ \ .
\label{eq:special}
\end{equation}
Hence we find
\begin{equation}
D(z_0)={\frac{c(cz_0+d)}{1-(cz_0+d)^2}} \ \ .
\label{eq:special3}
\end{equation}
The value of $D(z_0)$ of Eq.~(\ref{eq:special3}) determines {\it uniquely} 
the allowed values of the conductivities at the fixed points. However,
it is straightforward to check that
\begin{equation}
D(z)={\frac{i}{2 {\rm Im}z}}
\label{eq:special1}
\end{equation}
satisfies both the requirement of periodicity $P$ and it transforms correctly
under duality $S$. Hence, the value of $D(z)$ {\it at the fixed points} given 
by Eq.~(\ref{eq:special3}) {\it must} be equal to
\begin{equation}
D(z_0)={\frac{i}{2 {\rm Im}z_0}}
\label{eq:special4}
\end{equation}
Therfore, $D(z)$ is pure imaginary {\it at the fixed points}.
 
Let us derive its consequences for the specially important cases of bosons and 
fermions:
\newcounter{letters}
\begin{list}
{\alph{letters}~)}{\usecounter{letters}}
\item
Bosons: The boson fixed point $z_0=i$ is left invariant by the Duality
transformation, which is the $SL(2,Z)$ transformation with $a=d=0$ and
$b=-c=1$. Eq.~(\ref{eq:special3}) yields the value 
\begin{equation}
D(i)={\frac{i}{2}}
\label{eq:i}
\end{equation}
which is pure imaginary.
The conductivities for  a theory of self-dual  {\it bosons} become
\begin{eqnarray}
\sigma_{xx}&=&\sigma^0_{xx}-{\frac{e^2}{2 \pi}} d_S=\sigma^0_{xx}+
{\frac{e^2}{4 \pi}}\nonumber\\
\sigma_{xy}&=&\sigma^0_{xy}-{\frac{e^2}{2 \pi}} d_A=\sigma^0_{xy} \ \ .
\label{eq:sigma-bosons}
\end{eqnarray}
Notice that the odd-amplitude $d_A=0$. Except for the background conductivities 
$\sigma^0_{xx}$ and $\sigma^0_{xy}$, the conductivity $\sigma_{xx}$ of self-dual
 bosons is indeed universal and equal to ${\frac {1}{2}}{\frac {e^2}{2\pi}}$.
This is the conjectured value of the universal conductivity.

Likewise,  a generic {\it bosonic fixed point} ({\it i.~e.\/} a fixed point 
$z_0 \in [i]$)
has the form
\begin{equation}
z_0={\cal T}_{i \to z_0} \; i \equiv {\frac{ai+b}{ci+d}} \ \ .
\label{eq:fpi}
\end{equation}
Eq.~(\ref{eq:special4}) tells us that $D(z_0)$ is pure imaginary. The
longitudinal conductance at a generic boson fixed point is universal
although it is not equal to ${\frac{1}{2}} {\frac{e^2}{h}}$.
\item
Fermions: The fermion fixed point 
$z_0=\rho={\frac{1}{2}}+i {\frac{\sqrt{3}}{2}}$
is left invariant by Duality followed by Periodicty (with period 1). 
This is the modular transformation with $a=-b=c=1$ and $d=0$. 
For this point Eq.~(\ref{eq:special3}) yields the value
\begin{equation}
D(\rho)=i {\frac{\sqrt{3}}{3}}
\label{eq:rho}
\end{equation}
which, once again, is pure imaginary. 
The conductivities for self-dual  {\it fermions} are then given by
\begin{eqnarray}
\sigma_{xx}&=&
\sigma^0_{xx}-{\frac{e^2}{2 \pi}} d_S=\sigma^0_{xx}+
{\frac{e^2}{2 \pi}} {\frac{\sqrt 3}{3}}
\nonumber\\
\sigma_{xy}&=&\sigma^0_{xy}-{\frac{e^2}{2 \pi}} d_A=
\sigma^0_{xy} \ \ .
\label{eq:sigma-fermions}
\end{eqnarray}
Thus, the conductivities are again universal since,
 up to effects of the background conductivites, we get $\sigma_{xy}=0$ and 
$\sigma_{xx}={\protect{ {\frac{e^2}{2\pi}} {\frac{\sqrt{3}}{3}}}}$  as the 
value of the critical conductivity for this self-dual fermion system.
This analysis can be extended to all the {\it fermionic fixed
points}, $z_0 \in [\rho]$. Once again Eq.~(\ref{eq:special4}) 
yields a universal conductace.

It is interesting to consider the
fixed points (both bosonic and fermionic)  on the fermion line
$z={\frac{1}{2}}+ig$. These points are mapped 
into other points  on the same fermion line $z'={\frac{1}{2}}+{\frac{i}{4g}}$
 by the modular transformation $(a,b,c,d)=(-1,1,-2,1)$. This
transformation has one fixed point on this fermion line, namely the {\it
bosonic} point $z={\frac{1}{2}}+{\frac{i}{2}}$.
The transformation law of the amplitude $D(z)$ now becomes
\begin{equation}
D\left({\frac{1}{2}}+{\frac{i}{4g}}\right)=-4g^2 D\left({\frac{1}{2}}+ig\right)
+4ig \ \ .
\label{eq:transfermi}
\end{equation}
It is easy to check that $D\left({\frac{1}{2}}+{\frac{i}{2}}\right)=i$. 
Likewise, $z=\rho$ maps onto $z={\frac{1}{2}}+i{\frac{{\sqrt{3}}}{6}}$.
The transformation law now yields
$D\left({\frac{1}{2}}+i{\frac{{\sqrt{3}}}{6}}\right)=i{\sqrt{3}}$. 
Once again, we find a {\it universal} conductance.
\item
Extreme Fixed Points: These are the fixed points in the class
$[\infty] \cup \{ i \infty \}$. It is straightforward to show that $D(z)$ vanishes near
$\{ i \infty \}$. For instance, for $z=f+ig$ and  $g \to \infty$, the
contribution of the loops to the partition function is suppressed. In
the absence of loops the loop correlation functions vanish. 
More specifically, since the model is periodic in the statistical angle $f$, 
the partition function can be written as a sum over linking numbers or, what is the 
same, as the Fourier series
\begin{equation}
Z(f,g)=\sum_{\nu_L \in Z} Z_{\nu_L}(g) \; e^{2\pi i f \nu_L }
\label{eq:fourier}
\end{equation}
Here, the integer $\nu_L$ is the linking number and $Z_{\nu_L}(g)$ is the
partition function at fixed linking number. $Z_{\nu_L}(g)$ is positive and
it is a function of $g$ alone. 
For large but finite $g$, each term of this sum is of the order of 
$ \exp (-g S_0)$. The the action $S_0$ depends on the shape of the loops but it is
finite for large loops. Hence, for large $g$, only small loops will contribute
to $Z_{\nu_L}(g)$. In this regime, this series is convergent and, in fact
the partition function at statistics $f$ is bounded from above by the partition function 
for bosons (which also has an exponential dependence in $g$, at large $g$). 
Intuitively, this argument seemingly also implies that the leading non-zero contributions to 
the correlation functions are also exponentially small in $g$. If this were so, 
in this regime $D(z)$ would have an essential singularity in $g$, for large $g$.
While this is perfectly correct for a system of bosons, it is also conceivable that 
for generic statistics $f$ the phase fluctuations could conspire to
modify this behavior even at large $g$. In fact this is what should happen if 
Eq.~(\ref{eq:special4}) would hold even at large values of $g$  and not just at the fixed points. 
Without further elaboration we conclude that, in any case, we should expect $D(z)$ to vanish at 
large ${\rm Im}z$, either eponentially fast or with a weaker power law dependence.
In all scenarios, the physics of this system near $\{ i \infty \}$ is clearly that of an 
insulator.

On the other hand, near the other extreme fixed points $[\infty] $
(on the real axis $g \to 0$), the loops proliferate. However, since near these points 
the dual loops are suppressed, the behavior of the system in these regimes 
can be determined by using the transformation law. For instance, Duality
maps the {\it boson} points $z=ig$ onto the boson $z=-i/g$. 
For these points, the transformation law reads
\begin{equation}
D({\frac{i}{g}})=-g^2 D(ig)+ig  \ \ .
\label{eq:transbose}
\end{equation}
Hence, if $g \to 0$, this requires the {\it divergent} behavior  
$D(ig) \approx i/g$.
Similarly, the points on the fermion line $z={\frac{1}{2}}+ig$ for small
$g$ are mapped by the modular transformation $(-1,1,-2,1)$
onto points on the same line but now with large values of $g$. 
Finally, since in the extreme limit $g \to 0$ the long range interactions
effectively disappear, the behavior of the system is dominated by short
range interactions and/or mass terms. In particular, the free
Dirac fermion point $g \to 0$ has a Parity Anomaly whose precise value is
determined by the sign of the (infinitesimal) mass $m$ term~\cite{jackiw}.
In our language, the Parity Anomaly is the finite real limit   
$D(z={\frac{1}{2}}+i0^+)={\rm sign}(m) {\frac{1}{2}}$. We will not
elaborate on this interesting issue here any further. In any event,
even along the fermion line we expect essential singularities for large
$g$ (up to effects of the Parity Anomaly). Notice, however that since
the modular fixed points accumulate on the real axis, there will always
be a lenght scale in which the long range interaction dominate no matter
how small $g$ may be.
\end{list}

The analysis we have just presented shows that, {\it at least} for the
fixed points, $D(z)$ is completely determined and its values at the
fixed points are universal and independent of the analytic properties
that we may assign to it. Also, and in the same vein as in the
discussion of the Introduction, the fact that $\sigma_{xx}$ is neither
zero nor infinite in this system (which is not disordered) implies that,
at the modular fixed points, the system must necessarily be critical. 
This argument suggests very strongly that the system must also be scale
invariant at these fixed points.

It is natural to ask if these are the only points of the upper-half
plane $\cal H$ where $\sigma_{xx}\not= 0$, or, equivalently,
 ${\rm Im} D(z) \not= 0$. However, now we have to confront the problem that the
transformation law of Eq.~\ref{eq:G2} {\it alone} does not determine completely 
$D(z)$ any longer. To make further progress we need to know something about the
analytic properties of $D(z)$.

We have already argued above that if a fractional spin is not assigned
to the particles, this system has a time reversal invariance  which
requires the reflection symmetry of the amplitude $D(z)$. In turn,
reflection symmetry implies that $D(z)$ cannot be an analytic function
of $z$ in the sense that it is a function of both $z$ and its complex
conjugate $z^*$. It is natural to assume that there {\it could} be a
spin-statistics connection for which $D(z)$ {\it is} analytic. It is
also tempting to speculate that $D(z)$ {\it would} become analytic precisely 
for the {\it natural} spin-statistics connection (although we have no proof of 
this).

Let us assume momentarily that $D(z)$ {\it is} analytic ( holomorphic)
and derive the consquences that follow from this assumption. It
turns out that, on the upper-half plane $\cal H$, it is impossible to construct
a holomorphic function of $D(z)$ which transform {\it homogeneously} under
$SL(2,Z)$ like a modular form of weight $2$ since this would violate the Residue 
Theorem. However, in the theory of modular forms~\cite{modular,rankin}
it is proven that instead there exists a {\it unique} holomorphic function 
which, remarkably, transforms precisely as required by our inhomogeneous 
transformation law of Eq.~(\ref{eq:G2}). Thus, even though Model II may
not {\it require} analiticity it is certainly {\it compatible} with it.
The precise statement of this result is that $D(z)$ is a holomorphic
function of $z$ (hence, {\it only} a function of $z$) of the form
\begin{equation}
D(z)=-4 \pi i G_2(z) \ \ .
\label{eq:Dg2}
\end{equation}
Here $G_2(z)$ is the holomorphic function defined by the Fourier series
\begin{equation}
G_2(z)=-{\frac{1}{24}}+\sum_{n=1}^{\infty} \sigma_1(n) e^{2\pi i n z} \ \ ,
\label{eq:g2-fourier}
\end{equation}
where $\sigma_1(n)$ is the sum of all the divisors of the positive
integer $n$,
\begin{equation}
\sigma_1(n)=\sum_{r|n} r \ \ .
\label{eq:divisors}
\end{equation}
In terms of the variable $q=\exp (2\pi i z)$, the first terms of this 
Fourier expansion are
\begin{equation}
G_2(z)=-{\frac{1}{24}}+q+ 3q^2+4q^3+ \ldots \ \ .
\label{eq:terms}
\end{equation}
Alternatively, $G_2(z)$ can be written as a regularized Eiesenstein series
of the form
\begin{equation}
G_2(z)=- {\frac{1}{8\pi^2}} \lim_{\epsilon \to 0^+} \sum_{m,n} \;'
{\frac{1}{(mz+n)^2 |mz+n|^{\epsilon}}} + 
{\frac{1}{8 \pi {\rm Im} z}}
\label{eq:reg-g2}
\end{equation}
The significance of Eq.~(\ref{eq:Dg2}) is that, if analyticity were to 
hold exactly, the complex conductance would
be a {\it universal} function of $z=f+ig$, completely 
determined on the entire upper-half plane. 

In contrast, if analyticity does not hold, many options are
available. Indeed,  we have already noticed that
\begin{equation}
D(z)={\frac{i}{2 {\rm Im}z}}
\label{eq:special2}
\end{equation}
is a solution.  It also happens to satisfy reflection symmetry and it is 
also consistent with all the values of $D(z)$ that we have determined so far,
including the asymptotic behavior for $g \to 0$. However, we now have 
infinitely many solutions available to us. Indeed, let $F(z)$ be a
modular invariant (not necessarily holomorphic) function of $z$. An example of
such functions is the Jacobi Invariant $J(z)$ (which has an essential
singularity at infinity). The derivative $F'(z)$ is modular of weight $2$ 
(but not holomorphic). Then we have
\begin{equation}
D(z)={\frac{i}{2 {\rm Im}z}}+F'(z) \ \ .
\label{eq:general}
\end{equation}
For reflection symmetry to hold it is sufficient to write
\begin{equation}
D(z)={\frac{i}{2 {\rm Im}z}}+i |F'(z)| \ \ ,
\label{eq:general2}
\end{equation}
which satisfies all the requirements. 

Thus, analyticity is a very strong (and highly desirable!) constraint which determines 
uniquely the amplitude $D(z)$ and the conductances for all points of the upper 
half-plane. Analyticity appears to be related to the existence of a spin-statistics 
connection in a manner which is not yet fully understood. If analyticity
holds, our results show that it is possible to have a more general (and
interesting) type of superuniversality: the conductances are a universal
function of the coupling constant $g$ and of the statistical angle $f$.
In a sense, the complex conductance behaves very much like a universal
amplitude in critical phenomena, which are universal functions of
effective coupling constants. The latter, of course, are related in a
highly non-universal manner to the coupling constants that define the
physical systems at microscopic scales. The same structure appears to
hold here. We close by stressing that, even if analyticity does not hold,
modular invariance alone is sufficient to prove that the complex
conductance is different, but determined, at the different modular fixed points.

\section{Periodicity in QFT and Critical Behavior}
\label{sec:QFT}

In previous sections we derived a number of duality transformations for two classes of
theories of interacting loops endowed with statistical phases.

We showed that, for Model I and for a subset of possible values of the statistical
angle $f$, the theory
with short range interactions among loops is dual to a theory with long range,
logarithmic, interactions among loops. The theory with short range interactions among 
loops is essentially identical to the high temperature phase of the 3D $XY$ classical
Model. In the high temperature phase, the loops are the loops of the diagrams of the
high temperature expansion. These diagrams are also a lattice version of the Feynaman
diagrams of a theory of a self-interacting complex field. Close to the critical
coupling constant (or $T_c$) this theory becomes equivalent to $|\phi|^4$ theory. This
picture is known to hold from detailed studies of critical behavior  both within the
$4-\epsilon$ expansion and directly in three dimensions~\cite{zinn}.
 The dual theory is the low temperature phase of the 3D $XY$-Model. In this
regime the loops are  vortex defects of the order parameter field. The fact that a
theory with long range interactions among loops is dual to a theory with short range
loop interactions, both separated by the same second order phase transition, indicates
that the long range field of the vortices of the low temperature phase is actually
screened and that the actual fixed point is that of the $|\phi|^4$ theory. The
critical behavior of this theory is controlled by the Wilson-Fisher fixed point of its  
$4-\epsilon$ expansion. 
It is well established that the phase transition of the
3D $XY$ model is second order. The disordered phase can be viewed as a vortex 
condensate  but, unlike the 2D $XY$ model, it is not a
vortex-liberation phase transition. In fact, as the transition is approached, core
size of the core of the vortices grows large and the long
range interaction of the vortices becomes screened. The transition of the 3D $XY$
model is actually controlled by the Wilson-Fisher fixed point of $|\phi|^4$ Landau
theory. For other values of the statistical
angle $f$, the theory is equivalent to a $|\phi|^4$ theory coupled to a
Chern-Simons gauge field with a suitable value of the Chern-Simons coupling constant.

If matter is second-quantized, 
the situation is somewhat less clear. For second-quantized theories defined on a 
lattice~\cite{frohlich,ef,e-s,dst}, Periodicity always holds and so does Charge 
Conjugation in the appropriate cases. For non-relativistic dense systems in 
two-dimensional continuum space ({\it i.~e.~\/} all Chern-Simons theories of the FQHE) 
are invariant under $P$ and invariance under $C$ only happens under special 
circumstances. For continuum relativistic field
theories the situation is less clear since the symmetries depend on subtle properties
of this theory at short distances and are thus strongly sensitive to regularization
prescriptions. For instance, field  theories with a
single relativistic fermi field with a four-fermi interaction and coupled to a 
Chern-Simons gauge theory
have been considered recently by  several authors~\cite{wen-wu,matthew,weichen}.
Within a perturbative approach, which uses dimensional regularization, these 
authors study the critical behavior of such theories.
In particular, it is found~\cite{weichen}
that the dimension of the four-fermi operator $\Delta_4$ is a {\it continuous} 
function of the statistical angle $\delta$ (related to the Chern-Simons coupling
constant $\theta=1/2\delta$) and that, in particular, $\Delta_4$ {\it decreases} like 
$\Delta_4-4 \sim \delta^2$ (here $4$ is the dimension of the four-fermi operator at
the free field fixed point). Moreover, due to the topological charcter of the 
Chern-Simons action, its coupling constant acquires at most finite
renormalizations~\cite{semenoff-sodano-wu} and, consequently, it does not flow under
the renormalization group. Thus, if taken at face value, these results seem to imply 
that the critical properties of at least relativistic matter fields coupled to 
Chern-Simons gauge fields do depend parametrically on the value of the Chern-Simons 
coupling constant $\theta$ and that these transitions are neither necessarily 
superuniversal nor periodic in $\delta$. 

The results of references ~\cite{wen-wu,matthew,weichen} hold in a perturbative 
expansion in powers of $\delta$ ({\it i.~e.~\/} the deviation from fermions). 
Hence, periodicity cannot possibly be respected even if the full 
theory had this symmetry. However, the use of dimensional regularization imposes more 
serious limitations.
As a regularization procedure, dimensional regularization yields unambiguous results
only for theories (and observables) which can be continued in dimension in a natural
way. However, dimensional regularization is notoriously unreliable for the study of
theories with operators which involve the Levi-Civita epsilon tensor, such as 
Chern-Simons gauge theory itself and many others. The dimensional continuation of 
operators with the epsilon tensor is plagued with ambiguities with additional 
(or evanescent) operators which make important contributions even though they formally 
do not exist in three dimensions. From a physical point of view, the vanishing of all 
divergent contributions other than logarithmic, which in standard problems is an 
advantage of dimensional regularization, turns into a problem in this case. For
instance, in these theories there are ultraviolet linearly divergent  contributions to 
the fermion self-energy which induce a non-zero mass for the fermion. 
Since fermion masses break parity and time reversal in $2+1$-dimensions, they alter 
the symmetries of the system. Thus, different choices of cutoff may affect the physics.
For instance, in reference~\cite{heisenberg} the  anisotropic quantum Heisenberg
antiferromagnet was studied using a lattice Chern-Simons approach. This theory 
is a lattice version of the relativistic theory. It was found ~\cite{heisenberg} that
the parity breaking induced fermion mass terms are essential for the spectrum of the
theory to agree with the spectrum rigurously known for the anisotropic quantum 
Heisenberg antiferromagnet.

In any event, a full non-perurbative construction of these continuum
field theories in which Periodicity is an explicit symmetry is still
missing. Our discussion in the rest of the paper suggests a possible way
to do this. Our results indicate that perturbation theory may be
misleading.

\section{Conclusions}
\label{sec:conc}

In this paper we constructed of a set of models which
qualitatively describe the transitions between quantum Hall plateaus and
related problems. We gave an explicit construction of a model (Model II)
which is invariant under periodic shifts of the statistical angle and
under duality transformations. In fact the model is invariant under the
full Modular Group $SL(2,Z)$. We also investigated the transformation
laws under $SL(2,Z)$ of physical correlation functions. In particular we
showed that the complex conductance transforms (loosely speaking) like a
modular form of weight $2$. This observation alone enabled us to
determine the values of the complex conductance at the fixed points of
the modular group. The values of the conductances at these fixed points are
universal but not constant. 
We present an argument that shows that the system is critical at least at these 
points. In a sense the conductance behaves like a universal amplitude of a
critical system. Further assumptions about symmetries have to be made in order
to determine the conductance away from the fixed points. If the complex
conductance is a holomorphic function of $z=f+ig$, then the modular
transformation laws determine the conductance completely on the entire
upper half of the complex plane. This is perhaps the strongest form of
superuniversality. In fact, all the non-universal physics is
encapsulated in the dependence of the (long distance) effective coupling
constant $g$ on the microscopic coupling constants. For Time Reversal
Invariant systems we find that symmetry arguments alone are no longer
sufficient to determine the conductance. It is quite possible that the
system should have a hidden supersymmetry for $D(z)$ to be holomorphic.

A number of additional issues remain to be resolved. 
The models were constructed in such a way that particle-hole symmetry is exact. 
Also, we have assumed that there was no external magnetic field (on
average).
Clearly, in order  to describe the full set of phase transitions of the
FQHE these assumptions need to be removed. In the context of Model I
this has already been done in all of the studies of the FQHE plateaus.
However, for Model II we have chosen not to consider this generalization
here. Although in principle this is straightforward, this generalization
already poses an important question. In the context of Model II we have
shown that at the modular fixed points that complex conductance is pure
imaginary and this implies the vanishing of the  Hall conductance in a
modular invariant system. This result followed from the fact that the
complex conductance transforms like a modular form of weight $2$. We
have also observed that, in the presence of an  electromagnetic field,
the partition function is no longer a modular invariant. At finite fields
and chemical potential, the partition function will transform in a
non-trivial way. Thus, away from perfect particle-hole symmetry, the
system can have a non zero value of $\sigma_{xy}$ at the fixed points.
A related issue is the description in this language of the  Parity Anomaly 
in theories of relativistic fermions.

Finally, there is the question of the unitarity of Model II.
Since Model I describes a theory of particles with fractional statistics
with short range interactions it can also be written a {\it local} field
theory of a complex scalar field coupled to  a Chern-Simons gauge field.
Although the phase transitions of these theories are not understood,
their behavior at their stable, semiclassical regimes is well known. In
particular, for rational values of the statistical angle, these theories
are unitary. In contrast Model II, which turned out to be a more
interesting and richer theory, cannot be mapped onto a local field
theory for generic values of the statistical angle. Nevertheless, on any
bosonic line, the partition function has positive weights and it is
unitary. However, on any fermionic line $f={\frac{1}{2}}+n$ (with $n \in Z$)
Model II is also unitary since, on such lines the theory can be
related to a non-local version of a Maxwell-Chern-Simons theory. Since
Model II is modular invariant, it follows that unitarity should also hold
on the images of all of the bosonic and fermionic lines under the full action 
of $SL(2,Z)$. Therefore, we expect unitarity to hold on a dense set of points 
of the upper-half plane. However, it is quite possible (and likely) that 
unitarity may not hold for irrational values of the statistics.


\section{Acknowledgements}
\label{sec:ack}

We thank L.~Chayes, S.~Chaudhuri, D.~E.~Freed, D.~H.~Lee, C.~L{\"u}tken, D.~Rokhsar 
and M.~Stone for useful discussions. We are also very grateful to L.~Pryadko and 
S.~C.~Zhang for making their unpublished work available to us. 
This work was supported in part
by the National Science Foundation through the grants NSF DMR94-24511 at
the University of Illinois at Urbana-Champaign and NSF DMR-93-12606 at UCLA.


\begin{references}
\bibitem{spin} It is well known that, for a suitable choice of regularization, the 
Chern-Simons theory also describes fractional spin~\cite{polyakov}. 
However, this property does depend
on the microscopic definition of the theory and it is not universal. For the rest of
this paper, the effects of such a fractional spin will be ignored.
\bibitem{electro} In a physically relevant model which can describe a
condensed matter system, the velocity of the particles is much smaller
than the speed of light. A consequence of this is both an anisotropy of
the effective interaction and a significant suppression of the coupling
constant for the magnetic terms. In this regime, short range
interactions (in addition to the instantanoeus Coulomb piece) become
dominant. However, below we show that the behavior is highly non-analytic
at small coupling. Thus, even there, the magnetic terms are still important even if 
they are small.
\bibitem{foot} For simplicity, we fix the gauge by the standard 
method of adding a gauge fixing term to the action of the form 
${\frac{\alpha}{2}}\sum_{\vec R}[\Delta_\mu \phi_\mu(\vec R)]^2$. The final
expressions, themselves being gauge invariant, are independent of $\alpha$.
\end{references}
\end{document}